\documentclass[letterpaper, 10 pt, conference]{ieeeconf}  % Comment this line out if you need a4paper

\IEEEoverridecommandlockouts                              % This command is only needed if 
\usepackage{cite}
\usepackage{mathtools}
\usepackage{amsmath,amssymb,amsfonts}
\usepackage[english]{babel}
\usepackage{balance}
\usepackage{graphicx}
\usepackage{textcomp}
\usepackage{xcolor}
\usepackage{array}
\usepackage{tabularx}
\usepackage{multirow}
\usepackage{longtable}
\usepackage{algorithm}
\usepackage{algorithmic}
\usepackage{color, soul}

\newtheorem{theorem}{Theorem}[section]

\newtheorem{proposition}{Proposition}

\title{Defense against Joint Poisoning and Evasion Attacks: A Case Study of DERMS\LARGE \bf

\thanks{This project is funded by Commonwealth Cyber Initiative (CCI), C3.ai Digital Transformation Institute, and U.S. Department of Energy.}}%

\author{Zain ul Abdeen$^{a,*}$, Padmaksha Roy$^{a,*}$, Ahmad Al-Tawaha$^{a,*}$, Rouxi~Jia$^a$, Laura Freeman$^b$, Peter~Beling$^b$, Chen-Ching Liu$^a$, \\ Alberto Sangiovanni-Vincentelli$^c$, and Ming Jin$^a$}

\begin{document}
\maketitle
\thispagestyle{empty}
\pagestyle{empty}

%%%%%%%%%%%%%%%%%%%%%%%%%%%%%%%%%%%%%%%%%%%%%%%%%%%%%%%%%%%%%%%%%%%%%%%%%%%%%%%%
\begin{abstract}
There is an upward trend of deploying distributed energy resource management systems (DERMS) to
control modern power grids. However, DERMS controller communication lines are vulnerable to cyberattacks that could potentially impact operational reliability. While a data-driven intrusion detection system (IDS) can potentially thwart attacks during deployment, also known as  the evasion attack, the training of the detection algorithm may be corrupted by adversarial data injected into the database, also known as the poisoning attack.  In this paper, we propose the \emph{first} framework of IDS that is robust against joint poisoning and evasion attacks. We formulate the defense mechanism as a bilevel optimization, where the inner and outer levels deal with attacks that occur during training time and testing time, respectively. We verify the robustness of our method on the IEEE-13 bus feeder model against a diverse set of poisoning and evasion attack scenarios. The results indicate that our proposed method outperforms the baseline technique in terms of accuracy, precision, and recall for intrusion detection.
\end{abstract}
%\begin{keyword}cybersecurity of power systems, evasion attacks, poisoning attacks, intrusion detection systems, bilevel optimization\end{keyword}

%%%%%%%%%%%%%%%%%%%%%%%%%%%%%%%%%%%%%%%%%%%%%%%%%%%%%%%%%%%%%%%%%%%%%%%%%%%%%%%%
	\section{Introduction}
\label{sec:intro}
With the rapid digitization of societal-scale infrastructures, power systems are gradually being transformed into cyber-physical power systems (CPPSs), also known as smart grids. The use of distribution energy resources (DERs) such as rooftop photovoltaic and energy storage systems introduces variability in operations---uncontrolled variations in power injection can induce abrupt fluctuations in nodal voltages, jeopardizing system reliability \cite{r1}. Thus, distributed energy resource management systems (DERMS) are increasingly  deployed to manage the potential adverse impacts of DERs on distribution feeder voltages \cite{jain2021detection}. The centralized DERMS controller receives data streams from advanced metering infrastructure (AMI) and then decides upon optimal real and reactive power dispatch settings for inverter-based DERs \cite{dall2014optimal}. However, the heavy reliance on communications exposes the system to cyberattacks \cite{case2016analysis}. By targeting the DERMS communication channels, attackers can initiate falsified dispatch commands that cause severe voltage disturbances and damage substations or household equipment.
\begin{figure*}[t]
\begin{center}
\includegraphics[width=1.4\columnwidth]{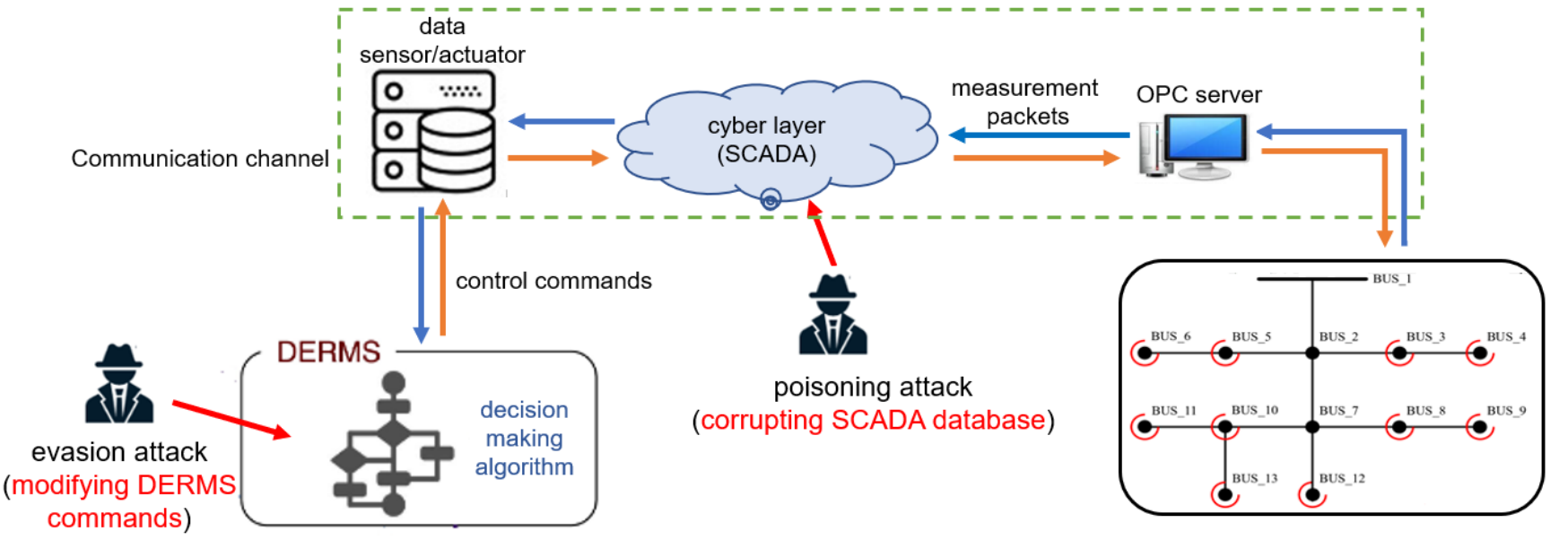}    % The printed column width is 8.4 cm.
\caption{Cyber-physical power system and its vulnerability. } 
\label{powermodel}
\end{center}
\end{figure*}
Cyber-vulnerability makes it imperative to study assessment and defense strategies \cite{ike2022scaphy}. The denial-of-service (DoS) attack \cite{chen2022distributed} and false data injection attack (FDIA) \cite{9244096} are two commonly analyzed attacks on the DERMS controller. Methods to detect and mitigate these attacks in the cyber-layer have also been investigated \cite{huseinovic2020survey,raja2022review}, including recent works with machine learning \cite{guo2021data,hasnat2021detecting,nguyen2021distributed}. Besides DoS and FDIA, a relatively low-probability but high-severity attack involves modifications to the DERMS controller algorithm after gaining unauthorized access \cite{jain2021detection}. As the software can be altered to disguise malicious command data packets, such attacks can be difficult to detect with a centralized method \cite{sun2021sok}. To counteract, decentralized inverter-based IDSs are proposed in \cite{jain2021detection,urbina2016limiting}; specifically, a regression model for expected control commands is trained with historical data and the prediction error is subsequently leveraged for evasion attack detection. Nevertheless, a crucial vulnerability persists: the historical data may be adversarially manipulated by a data poisoning attack; as a consequence, the trained model may trigger false alarms or miss attack events when deployed in test time \cite{tian2022comprehensive}. This calls for an IDS that is robust to attacks that may occur at different stages.

In this paper, we focus on the challenging scenario where both poisoning (training phase) and evasion (testing phase) attacks can be staged. For the development of the defense mechanism, our key insight is that as the trained model will be used subsequently for evasion attack detection, such model should be trained robustly and in an end-to-end fashion. The contributions are summarized as follows:
\begin{itemize}
    \item Development of a model-based IDS against joint poisoning and evasion attacks on DERMS;
    \item Formulation of a bilevel optimization problem, where the inner level robustly learns a model and the outer level finds an optimal threshold for the model-based prediction error;
    \item Evaluation of the proposed method in a range of attack scenarios and demonstration of improved robustness against the baseline method.
\end{itemize}

The rest of the paper is organized as follows. Section~ \ref{sec:related} presents related works on cyber-defense mechanisms for power grids. Section~\ref{sec:psam} discusses backgrounds on CPPS and the attack model. The defense strategy is presented in Section~\ref{sec:defense}. Section~\ref{sec:experiment} conducts numerical evaluation of the proposed method and discusses the results. Finally, Section~\ref{sec:conclude} concludes the work. 

\section{Related work}
\label{sec:related}
%Due to the security vulnerabilities of the smart grid and its direct impact on day-to-day lives, many smart grid security and attack detection methods have been proposed recently  \cite{tuyen2022comprehensive,zhang2021deep}.\cite{huseinovic2020survey,raja2022review},delay attacks \cite{ganesh2021learning}, man-in-the-middle (MITM) attacks \cite{osti_1798924}, replay attacks \cite{elsaeidy2020replay} and false data injection attacks (FDIAs) \cite{liang2016review,liu2016false}. 

Cyberattacks on power grids include DoS, delay attacks, man-in-the-middle attacks, replay attacks, FDIAs, just to name a few \cite{peng2019survey,tuyen2022comprehensive}. Various defense mechanisms have been proposed \cite{sun2018cyber,peng2019survey}, among which machine learning (ML) techniques are promising \cite{berghout2022machine}. Existing data-driven IDSs can be categorized into supervised learning \cite{wang2019detection}, semi-supervised learning \cite{farajzadeh2021adversarial}, unsupervised learning \cite{karimipour2019deep}, self-supervised learning \cite{zhang2021deep}, and reinforcement learning \cite{kurt2018online}. To improve data efficiency, model-based defense mechanisms can leverage physics and have been developed to detect evasion attacks \cite{karimipour2017robust,jain2021detection}. 

Nevertheless, the effectiveness of ML-based IDS can be significantly reduced by poisoning attacks, which have become an emerging threat \cite{tian2022comprehensive}. Different from an evasion attack that occurs during deployment (test time), a poisoning attack can misguide the model training by manipulating the training data, thus yielding a falsified model for deployment. Defending against poisoning attacks is challenging and less investigated for power system cybersecurity \cite{zografopoulos2022distributed}. Furthermore, very few works have addressed the scenario of joint poisoning and evasion attacks, leaving a substantial gap in the literature. The challenge is to reason about the propagation of error induced by poisoning attacks to test time performance and then design a learning framework that is robust to such error propagation. 

The present study initiates the first study in this important direction. Our technique hinges on bilevel optimization \cite{liu2021investigating}. The methodology design is inspired by the recent line of research on end-to-end optimization \cite{kotary2021end}, which provides a principled way to design the training of a model in view of its consequent usage during test time.
\section{Power system and attack model}
\label{sec:psam}
\subsection{Cyber-physical power system}

%\jin{Describe the physical system. DERMS controller. }
The physical layer of a CPPS consists of the feeders and DERMS controller and the cyber layer represents the information and communication technology (ICT) or the supervisory control and data acquisition (SCADA) system and the communication paths for data exchange (see Fig.~\ref{powermodel} for an illustration with the IEEE-13 bus feeder model) \cite{jain2021detection}. The open platform communication (OPC) server receives measurement packets from the feeder and prepares the data for SCADA to access in the cyber layer. Measurement data are sent to the DERMS controller through a firewall to check for discrepancies. The control actions, such as the optimal real and reactive power setpoints, are computed by the DERMS controller and sent back to the feeder for actuation. An IDS is deployed within the cyber layer to constantly check for the integrity of data and control commands.

\subsection{Attack model}
Due to the heavy reliance on ICT, the attack surface is wide in practice and may include data integrity injection at substations and over communication links, distributed attacks by manipulating endpoint devices such as smart meters and smart appliances, or even a more powerful attack such as spear phishing attacks that gain access to communication paths or modify the DERMS controller software \cite{adepu2018epic}. The goal of the attacker can be to falsify the dispatch control commands to cause voltage violations and exact damage to the physical systems. Nevertheless, among all the possibilities, some of the attacks can be more severe (e.g., taking   over control centers) than others (e.g., attacking smart appliances). As a consequence, the attack may range in severity due to the ability of the attacker. 

In this study, we consider two types of threats: evasion attacks and poisoning attacks. While these threats can be implemented with one or a combination of the aforementioned attacks, the key difference is the time when the attack is staged: an evasion attack occurs during deployment to evade IDS, whereas a poisoning attack may be conducted in an earlier stage during model training to corrupt the IDS. 

Our assumption of the attacker is comparable to the existing works on data integrity attacks \cite{liang2016review}; in particular, we assume that the data used for training or during actual operations can be maliciously manipulated. We remark that the attacker considered in our study is stronger than some existing works on model-based defense \cite{jain2021detection,ghaeini2018state} in the sense that prior works focus on evasion attacks while we consider the additional mode of poisoning attacks. This stronger attack model seems more practical due to the increasing use of ML in modern IDS and the various security loopholes in database systems \cite{tian2022comprehensive,ike2022scaphy} .

\section{Defense strategy}
\label{sec:defense}
\subsection{Decentralized detection}\label{sec4.1}
In model-based IDSs, the expected command is compared against the actual command received, and an anomaly is detected when the difference between these values is larger than a threshold \cite{9244096,jain2021detection}. As the cyberattack may directly target the DERMS software to make malicious data packets appear legitimate, a centralized IDS may be evaded, while a decentralized approach that uses locally available measurements may be more difficult to deceive. While our framework can incorporate more complex models such as neural networks \cite{9244096}, due to limited computational power at the inverter level, a simple model such as linear regression is preferred. Following \cite{jain2021detection,zeng2021adversarial}, a linear regression model is used to predict the expected control commands based on local load and maximum charging and discharging rate values. For instance, the expected control command for real power dispatch set point is given by 
\begin{equation}\label{ppred}
    P_{D}^{pred}= \alpha_{D}^{1,p}+\alpha^{2,p}_{D}*p_{L}+\alpha^{3,p}_{D}*q_{L}+\alpha^{4,p}_{D}*p_{Dmax },
\end{equation}

where $p_{L}$ and $q_{L}$ represent the active and reactive power demands, respectively, and $p_{Dmax}$ is the maximum generation limit of the inverter. Here, $\{\alpha_{D}^{j,p}\}_{j=1,\dots,4}$ are  coefficients of the regression model. Note that we can write \eqref{ppred} as $P_{D}^{pred}=\alpha\cdot x$, where $x=[ 1, p_{L},q_{L},p_{Dmax} ]^{\top}$ and $\alpha= [\alpha^{1,p}_{D},\alpha^{2,p}_{D},\alpha^{3,p}_{D},\alpha^{4,p}_{D}]^{\top}$.

In the following, we denote $x_{i}$ as the feature vector for data point $i$ (so $\alpha\cdot x_{i}$ is the expected command) and $p_{i}$ as the actual command. For each data point $\left(x_{i},p_{i}\right)$, if the absolute difference $|\alpha \cdot x_{i}-p_{i}|$ between the expected and actual commands is larger than a threshold $\tau$, then we consider that an anomaly has occurred; otherwise, the data point is considered normal. Similar models can be instantiated for other control commands, such as reactive power dispatch set points for PV inverters and charging or discharging rates for energy storage inverters. To streamline the presentation, we will focus on the real power dispatch set point for illustration.

\subsection{Bilevel formulation of defense}
\textbf{Problem setup.} Let $\mathcal{D}_{1}=\{(x_i,p_i)\}_{i=1}^{n_1}$ be an unlabeled dataset, where $x_{i}\in \mathbb{R}^{4}$ is the feature vector and $p_{i}$ is the actual command. Suppose we also have access to a dataset that contains labels regarding whether an attack has occurred, i.e., $\mathcal{D}_{2}=\{(x_i,p_i,y_i)\}_{i=1}^{n_2}$, where $y_{i}\in \{-1,+1\}$ is the label with $+1$ indicating the anomaly and $-1$ indicating the normal condition. In practice, as cyberattack data are rare and difficult to obtain, we expect that the amount of unlabeled data to be much larger than the amount of cyberattack data, namely $n_1\gg n_2$. Based on our attack model, a certain (but unknown) percentage of the dataset $\mathcal{D}_1$ may be poisoned; thus the actual measurements $p_i$ cannot be trusted. We assume that the labels $y_i$ in $\mathcal{D}_2$ are authentic, since they are often carefully cross-checked by experts, although it is possible to extend our framework to consider corrupted labels as well.

Due to the presence of poisoned data during training, the conventional pipeline that first estimates the model parameter $\alpha$ with $\mathcal{D}_1 \cup \mathcal{D}_2$ and then uses the learned model to detect evasion attacks may no longer be effective \cite{jain2021detection,9244096}. Our strategy is differentiated from prior works in two-fold: \emph{1)} the training algorithm to obtain $\alpha^{*}$ should be robust to poisoning attacks on $\mathcal{D}_1$, and \emph{2)} as $\alpha^*$ is used in the downstream decision task---evasion attack detection---so the search of the prediction model should be aware of this task. We address these two aspects as follows.

\textbf{Robust training against poisoning attacks.} To design a robust training algorithm, we formulate the following optimization problem:
\begin{equation}\label{eq2}
    \arg\min_{\alpha,\delta}\frac{1}{2}\sum_{(x_i,p_i)\in\mathcal{D}_{1}}(\alpha\cdot x_i -p_i +\delta_i)^2+\lambda\|\delta\|_1,
\end{equation}
where $\|\cdot\|_1$ is the $\ell_{1}$ norm, $\lambda$ is a hyperparameter, and $\delta =[\delta_1,\dots,\delta_{n_{1}}]^{\top}$ is hypothetical bad data vector, which is introduced to counterbalance the potential attacks on $p_{i}$. The training loss consists of two terms: the squared loss of reconstruction error and a penalty on the sparsity of $\delta$. The overall problem is convex; in fact, it is strongly convex due to the presence of the squared loss, thus the optimal solution is unique. Under certain conditions, it has been shown that we can exactly recover the poisoned data by solving \eqref{eq2} \cite{jin2020boundary}. However, it is difficult to determine the best way to set $\lambda$---a larger value may induce a sparser $\delta$ but also a higher loss on the reconstruction error, and vice versa. While prior works set this number by hand, we propose to set this number so that it supports the ultimate task assigned for the model: evasion attack detection.

\textbf{Task-aware learning for evasion attack detection.} The model parameter $\alpha$ is used to detect evasion attacks by checking the prediction error, which provides a proper goal to guide the search of hyperparameter $\lambda$. Furthermore, the detection threshold $\tau$ needs to be tuned to support this task. To this end, a bilevel optimization problem is formulated:
\begin{equation}\label{eq3}
\begin{aligned}
    & \underset{\lambda,\tau}{\min}\sum_{(x_i,y_i,p_i)\in\mathcal{D}_{2}}\ell(|\bar{\alpha}\cdot x_{i} - p_i|-\tau,y_i)\\ & \text{s. t.}\quad \left(\bar{\alpha},\delta^{*}\right)~\text{is the optimal solution to \eqref{eq2}}
     \end{aligned}
\end{equation}
where $\ell(t,y)=\log(1+\exp(-ty))$ is the logistic loss. In the above formulation, the inner level determines the model parameters $\bar{\alpha}$ and hypothetical bad data vector $\delta^{*}$, while the outer level determines the hyperparameter $\lambda$ used within the inner level and the detection threshold $\tau$. The optimal $\tau$ depends on the learned model $\bar{\alpha}$, which in turn depends on the hyperparameter $\lambda$. Since the inner-level problem variable is included in the upper-level problem, in the case of poisoning attacks, the attack error may propagate into the evasion attack performance. Thus, the outer level also plays a role in rectifying the learned model to ensure that $\tau$ properly accounts for the potential corrupted model. Above all, the outer-level has fewer decision parameters than the inner-level, which agrees with the imbalanced data sizes $(n_1 \gg n_2)$. 

\subsection{Algorithm}
 To solve \eqref{eq3}, we can use gradient descent on the outer-level variables, while solving the inner-level problem exactly in each iteration (see Algorithm \ref{alg:iterative_impl_grad1}). Let $L(\lambda,\tau)=\sum_{(x_i,y_i,p_i)\in\mathcal{D}_{2}}\ell(|\bar{\alpha}\cdot x_{i} - p_i|-\tau,y_i)$ denote the outer-level objective. The gradients of $L$ with respect to $\lambda$ and $\tau$ are given by:
 \begin{equation}\label{eq4}
    \frac{\partial L}{\partial \tau}=\sum_{(x_i,y_i,p_i)\in\mathcal{D}_{2}}\frac{y_i\exp(-y_i(|r_i|-\tau))}{1+\exp(-y_i(|r_i|-\tau))},
\end{equation}
and
\begin{equation}\label{eq5}
    \frac{\partial L}{\partial \lambda} =\sum_{(x_i,y_i,p_i)\in\mathcal{D}_{2}}\frac{-y_i\mathrm{sign}(r_i)\exp(-y_i(|r_i|-\tau))}{1+\exp(-y_i(|r_i|-\tau))}{x}_i^{\top}\frac{\partial \bar{\alpha}(\lambda)}{\partial \lambda},
\end{equation}
where $r_i =\bar{\alpha}\cdot {x_i}-p_i $ and $\mathrm{sign}(r_i)=1$ if $r_i\geq 0$ and $0$ otherwise. As $\bar{\alpha}$ is a function of $\lambda$, the key is to obtain the gradient $\frac{\partial \bar{\alpha}(\lambda)}{\partial \lambda}$.

\textbf{Implicit gradient.} The difficulty of obtaining the implicit gradient from the usual Karush--Kuhn--Tucker (KKT) conditions is due to the presence of $\|\cdot\|_{1}$, which is not differentiable. In the following, we provide a closed-form solution to the implicit gradient $\frac{\partial \bar{\alpha}(\lambda)}{\partial \lambda}$. We start with a proposition that reformulate the inner-level problem as an optimization over the Huber loss:
\begin{equation}\nonumber
    f_{\mathrm{Huber}}(z;\lambda)=\begin{cases}
        \frac{1}{2}z^2&|z|\leq\lambda\\
        \lambda(|z|-\frac{1}{2}\lambda)&|z|>\lambda
    \end{cases}.
\end{equation}

\begin{proposition}\label{prop1}
Suppose that $(\bar{\alpha},\delta^*)$ is the solution to \eqref{eq2} and let $\bar{\alpha}'$ be the solution to $\min_{\alpha}\sum_{(x_i,p_i)\in \mathcal{D}_1}f_{\text{Huber}}(\alpha\cdot x_{i}-p_{i};\lambda)$. Then, we have $\bar{\alpha}=\bar{\alpha}'$, and the $i$-th component of $\delta^*$ is given by:
\begin{equation}\nonumber
    \delta_i^*=\mathrm{sign}(p_i-\bar{\alpha}\cdot x_i)\max(0,|\bar{\alpha}\cdot x_i-p_i|-\lambda).
\end{equation}

\end{proposition}
 The implication of the above result is that we can eliminate the inner-level variable $\delta$ and exclusively focus on $\alpha$ by changing the loss function. The following result provides the closed-form solution to the implicit gradient.
 \begin{theorem}\label{th1}
 Suppose that $(\bar{\alpha},\delta^*)$ is the solution to \eqref{eq2}. Let $\mathcal{I}_{1}=\{i:|\bar{\alpha}\cdot x_{i}-p_i|<\lambda\}$, $\mathcal{I}_{2}=\{i:\bar{\alpha}\cdot x_{i}-p_{i}\leq -\lambda\}$, and $\mathcal{I}_{3}=\{i:\bar{\alpha}\cdot x_{i}-p_{i}\geq \lambda\}$ be partition of dataset $\mathcal{D}_{1}$. Additionally, let $A=\sum_{i\in \mathcal{I}_{1}}x_{i}x_{i}^{\top}\in \mathbb{R}^{d\times d}$ and suppose that $A$ is invertible. Then, we have that
 \begin{equation}
 \label{eq:dadlambda}
    \frac{\partial \bar{\alpha}}{\partial \lambda}=\lambda A^{-1}\left(\sum_{i\in\mathcal{I}_2}x_i-\sum_{i\in\mathcal{I}_3}x_i\right).
\end{equation}
 \end{theorem}

 \begin{algorithm}
 \caption{Bilevel optimization algorithm  for \eqref{eq3}  }
 \label{alg:iterative_impl_grad1}
 \begin{algorithmic}[1]
 \renewcommand{\algorithmicrequire}{\textbf{Input:}}
 \renewcommand{\algorithmicensure}{\textbf{Output:}}
 \REQUIRE stepsize $\beta_{\tau}$ and $\beta_{\lambda}$, iterations $K$, initial values of $\tau_{1}$ and $\lambda_{1}$
  \FOR { $k=1$, $\dots$, $K$}
  \STATE Solve the lower level problem \eqref{eq2} to obtain $\alpha_k$
   \STATE Obtain $\frac{\partial L}{\partial \tau}$ and $\frac{\partial L}{\partial \lambda}$ at $\tau_{k}$ and $\lambda_{k}$ using \ref{eq4} and \ref{eq5}, respectively
  \STATE Update the value of $\tau$ and $\lambda$ using gradient descent
    \begin{equation*}
    \tau_{k+1} = \tau_k - \beta_{\tau} \frac{\partial L}{\partial \tau}, 
\end{equation*}  
\begin{equation*}
    \lambda_{k+1} = \lambda_{k} - \beta_{\lambda} \frac{\partial L}{\partial \lambda}
\end{equation*}
  \ENDFOR
  \STATE{\textbf{Output:} $\tau_{K}$ and $\alpha_{K}$}

  \end{algorithmic}
 \end{algorithm}

\section{Numerical evaluation}
\label{sec:experiment}
\textbf{Experimental setup.} We follow the same procedure as \cite{jain2021detection} to obtain the datasets, with $n_1=1000$ for the unlabeled dataset $\mathcal{D}_{1}$ and $n_2=200$ for the labelled dataset $D_{2}$. In $\mathcal{D}_1$, we consider the cases where $10\%$ and $30\%$ of the data are poisoning attacked. Dataset $\mathcal{D}_2$ consists of $20\%$ of data with label $+1$, i.e., anomaly. We also vary the levels of corruption by changing the true measurement of $p_i$ by the percentages of $40\%$,$70\%$, and $100\%$, with random noises of small magnitudes added upon the obtained values. 

During the training stage, we solve $\eqref{eq3}$ with the datasets $\mathcal{D}_{1}$ and $\mathcal{D}_2$ to obtain the solution $({\alpha^{*}},\tau^{*})$. During testing, we use the decentralized detection method outlined in Sec.~\ref{sec4.1} to detect evasion attacks. We evaluate the performance of our method in terms of metrics including the accuracy, precision and recall. Specifically, let TP, TN, FP, and FN denote the true positives, true negatives, false positives, and false negatives, respectively. Then, we have that 
\begin{align*}
    \text{accuracy}&=\frac{\text{TP+TN}}{\text{TP+TN+FP+FN}},\\
    \text{precision}&=\frac{\text{TP}}{\text{TP+FP}},\quad\text{recall}=\frac{\text{TP}}{\text{TP+FN}}.
\end{align*}

\textbf{Baseline method.} As a comparison, we also implement the approach from \cite{jain2021detection}. To briefly recap their method, a model parameter $\bar{\alpha}$ is first learned by solving a standard least-square regression problem on $\mathcal{D}_1$. Then, the threshold $\tau$ is manually designed based on the obtained $\bar{\alpha}$. To make a fair comparison, we also fine-tune the threshold based on $\mathcal{D}_2$.

\textbf{Results and discussions.} Tables \ref{tab:case1} and \ref{tab:case2} show the performance metrics of our proposed approach and the baseline method corresponding to $10\%$ and $30\%$ of poisoning attacks on $\mathcal{D}_{1}$, respectively. We report the mean and the standard deviation over $10$ independent runs. In general, it can be observed that the proposed method has improved accuracy, precision, and recall compared to the baseline. The improvement is more substantial in the case where $30\%$ of $\mathcal{D}_{1}$ are poisoning attacked (Table \ref{tab:case2}). This is expected as the baseline method uses linear regression to learn the model, which is well-known to be vulnerable to outliers or adversarial data.

As the \emph{evasion attack magnitudes} increase from $40\%$ to $100\%$, an interesting trend can be observed that the performance of each method (ours and baseline) increases. This is because for attacks with larger magnitudes, the differences between the expected and actual commands may easily surpass the detection threshold, even if the prediction model is not reliable. 

As the \emph{poisoning attack magnitudes} increase from $40\%$ to $100\%$, there is a clear trend that the performance of the baseline method drops. In contrast, in many cases, we can actually observe a slight increase in performance as the poisoning attack magnitudes increase from $40\%$ to $70\%$. This benefits from the robust training procedure in the inner-level problem, which can more easily detect adversarial data with a large deviation from normal. However, as the attack magnitude further increases from $70\%$ to $100\%$, even a few undetected outliers may significantly bias the training outcome, thus we can see a slight decrease in performance in some cases.

For both methods, we observe a higher precision than the recall. This can be attributed to the fact that the datasets $\mathcal{D}_{1}$ and $\mathcal{D}_{2}$ have imbalanced labels---the amount of anomaly data is fewer that the amount of normal data. This is generally to be expected, as anomalies often occur rarely. It may be an interesting direction for future work to test methods such as cost-sensitive learning or X-risk optimization to optimize for compositional measures \cite{yang2022algorithmic}.

\begin{figure}
\centering
\includegraphics[width=0.5\textwidth]{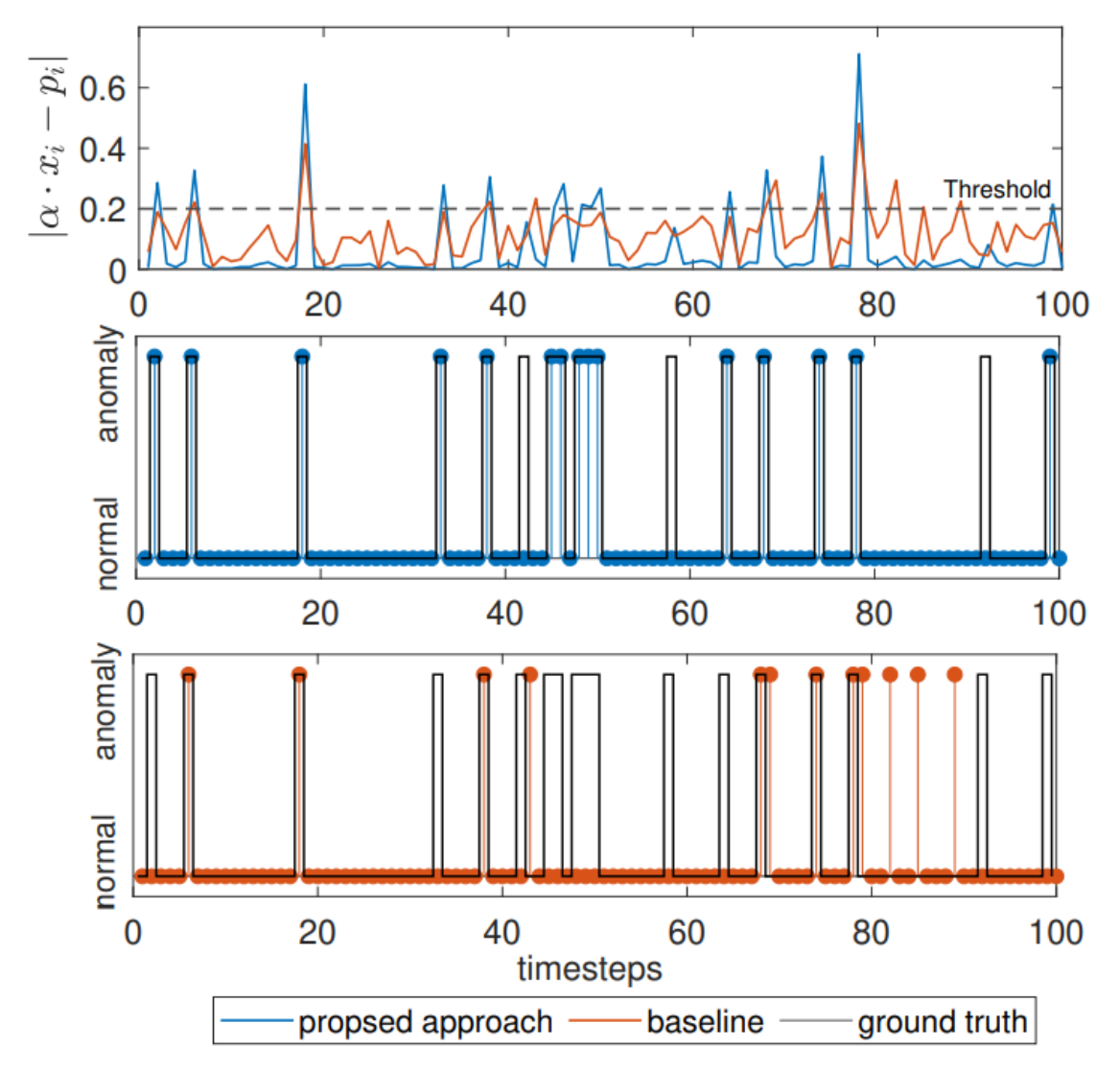}    
\caption{Visualization of the detection performance. Top plot: time series of the difference between predicted and actual commands (blue: ours, red: baseline). Middle/bottom plots: detection of anomaly based on the proposed method/baseline technique (stem plots). The ground truth is marked with the square wave. } 
\label{fig4}

\end{figure}

Last, we visualize the performance of the IDSs in a case with $70\%$ and $100\%$ attack magnitudes by evasion and poisoning attacks, respectively, as shown in Fig \ref{fig4}. As shown in the top figure that plots the difference between actual and expected commands, there are many instances of disagreements between our method and the baseline. Further examinations in the bottom two subplots indicate that in most cases, our method is able to accurately detect anomalies while avoiding false positives. In contrast, the baseline methods create multiple instances of false positives and false negatives, due to the corrupted prediction model affected by the poisoning attacks. 

\begin{table*}[t]
\resizebox{2\columnwidth}{!}{
\begin{tabular}{lcccccccccc}
\hline {} & {} &  \multicolumn{3}{c}{ 40\% evasion attack magnitude } & \multicolumn{3}{c}{ 70\% evasion attack magnitude}& \multicolumn{3}{c}{100\% evasion attack magnitude}   \\
\cline{3-5} 
\cline{6-8}
\cline{9-11}

 method & poisoning  & accuracy & precision & recall & accuracy & precision & recall & accuracy & precision  & recall \\
% \cline { 3 - 4 } \cline { 6 - 8 } \cline { 10 - 12 } & Method & Poisoning  && Accuracy & F1-Score & Recall && Accuracy & F1-Score & Recall && Accuracy & F1-Score & Recall \\

\hline proposed & $40\%$ & $\mathbf{86.4(2.4)}$ & $\mathbf{93.9(1.25)}$ & $\mathbf{59.3(3.5)}$ &

$\mathbf{93.8(3.8)}$ &
$\mathbf{96.3(2.2)}$&
$\mathbf{85.5(6.0)}$&

$\mathbf{93.1(2.5)}$&
$\mathbf{96.0(1.5)}$&
$\mathbf{82.1(5.7)}$\\

approach & 
$70\%$ & $\mathbf{84.4(3.7)}$ & $\mathbf{91.8(1.8)}$ & $\mathbf{60.7(4.3)}$ &

$\mathbf{94.5(2.5)}$ & $\mathbf{96.8(1.4)}$& $\mathbf{85.0(7.1)}$ &

$\mathbf{95.4 (1.7)}$ & $\mathbf{97.3(1.0)}$ & 
$\mathbf{88.0(3.7)}$ \\

& $100\%$ & $\mathbf{85.1(3.1)}$ & $\mathbf{92.2(1.7)} $ & $\mathbf{61.5(3.1)} $

& $\mathbf{92.6 (1.8)} $ & $\mathbf{95.8(1.0)}$  & $\mathbf{79.5(5.1)} $ &

$\mathbf{94.6(2.0)}$ & $\mathbf{96.8(1.1)}$ & $\mathbf{86.9(5.7)}$\\

\hline baseline & $40\%$ & $85.3(2.5)$ & $86.2(16.7)$ & $55.8(4.2)$ &

$93.6(3.9)$ & $96.3(2.2)$ & $85.1(5.5) $ &

$92.2 (3.3)$ & $95.6(1.9)$ & $80.0(6.5)$ \\

approach & $70\%$ & $82.9(3.5)$ & $78.7(22.0)$ & $56.7(5.3)$ &

$93.8(2.1)$ & $96.5(1.4)$ & $83.6(6.0)$ & 

$95.0(1.4)$ & $97.1(1.0)$ & $87.0(2.8)$ \\

& $100\%$ & $82.5(3.2)$ & $78.6(21.6)$ & $54.4(3.2)$ &

$91.1(1.3)$ & $95.1(1.0)$ & $79.5(5.2)$ & 

$93.4(2.1)$ & $95.7(2.0)$ & $84.4(4.9)$ \\
\hline
\end{tabular} }
 \caption{\label{tab:case1}Performance of the proposed method and baseline method with 10\% of dataset $\mathcal{D}_1$ under poisoning attacks. Each row indicate the performance of the corresponding method when the training data is poison attacked with magnitude 40\%, 70\%, or 100\%. The mean and the standard deviation (in paranthesis) are reported  over 10 independent runs. We mark better performance measures by boldface.}
\end{table*}

\begin{table*}[t]
\resizebox{2\columnwidth}{!}{
\begin{tabular}{lcccccccccc}
\hline {} & {} &  \multicolumn{3}{c}{ 40\% evasion attack magnitude } & \multicolumn{3}{c}{ 70\% evasion attack magnitude}& \multicolumn{3}{c}{100\% evasion attack magnitude}   \\
\cline{3-5} 
\cline{6-8}
\cline{9-11}
method & poisoning & accuracy & precision & recall & accuracy & precision & recall & accuracy & precision & recall \\
\hline proposed & $40\%$ & $\mathbf{82.8(3.4)}$ & $\mathbf{91.06(1.8)}$ & $\mathbf{58.8(4.3)}$ &

$\mathbf{94.1(2.2)}$ &
$\mathbf{96.6(1.3)}$&
$\mathbf{83.7(4.1)}$&

$\mathbf{94.0 (2.8)} $&
$\mathbf{96.5(1.7)}$&
$\mathbf{84.7(5.3)}$\\
approach & 
$70\%$ & 
$\mathbf{82.6 (4.1)} $ &
$\mathbf{90.8(2.1)}$ & 
$\mathbf{61.0(3.4)}$ &

$\mathbf{95(1.31)}$ & 
$\mathbf{97.1(0.7)}$ &
$\mathbf{85.6(3.6)}$ &

$\mathbf{92.0(3.5)}$ &
$\mathbf{95.3(2.1)}$ & 
$\mathbf{82.1(4.8)}$ \\

& $100\%$ & $\mathbf{83.6(2.4)}$ & $\mathbf{91.3(1.2)}$ & 
$\mathbf{63.2(3.2)}$ &

$\mathbf{93.4(2.3)}$ & 
$\mathbf{96.2(1.2)}$  &
$\mathbf{83.2(6.3)}$ &

$\mathbf{92.9(2.5)}$ &
$\mathbf{95.9(1.4)}$ & $\mathbf{81.6(6.5)}$\\

\hline baseline & $40\%$ & $81.5(3.6)$ & $84.3(16.5)$ & $55.6(3.4)$ &

$91.8 (3.6) $ & $95.4(2.0)$ & $78.0(5.6)$ & 

$93.7 (3.0)$ & $96.4(1.8)$ & $84.1(5.2)$ \\

approach & $70\%$ & $75.3 (5.4)$ & $48.1(18.5)$ & $48.8(2.2)$ &

$87.5(4.2)$ & $79.8(11.2)$ & $73.4(8.5)$ & 

$89.2(4.0)$ & $91.8(4.7)$ & $76.9(5.4)$ \\

& $100\%$ & $71.1(3.6)$ & $44.5(2.4)$ & $47.3(1.6)$ & 

$76.9(9.9)$ & $65.3(11.7)$ & $59.0(10.1) $ & 

$83.5(4.9)$ & $75.3(9.2)$ & $69.4(8.2)$ \\
\hline
\end{tabular} }
\caption{\label{tab:case2}Performance of the proposed method and baseline method with 30\% of dataset $\mathcal{D}_1$ under poisoning attacks. See Table \ref{tab:case1} for other descriptions.}
\end{table*}

\section{Conclusion}
\label{sec:conclude}

Cybersecurity is a tug of war---as the attacker's capability grows, so should the defender's, and it has strategic value in assuming a strong attacker so the defense can be assessed and designed commensurately. In this paper, we envisioned joint poisoning and evasion attacks on both the cyber and physical layers of a power system, a scenario that has not been systematically studied in the literature. As a countermeasure, we design a defense mechanism by formulating a bilevel optimization problem, where the inner level and outer level work jointly but with different goals. In particular, the inner-level problem accounts for robust training against poisoning attacks, whereas the outer-level problem addresses evasion attacks by guiding the inner-level training through implicit gradients.  The robustness of the method is evaluated under different attack scenarios and compared with a baseline model. In the future, we plan to evaluate our model on a real-time  digital simulator for power systems. Another interesting direction is to learn a nonlinear model through convexification to account for AC power flows \cite{jin2018power}.

%%%%%%%%%%%%%%%%%%%%%%%%%%%%%%%%%%%%%%%%%%%%%%%%%%%%%%%%%%%

  \appendix  
\section{Proofs}
\subsection{Proof of Proposition \ref{prop1}}
Given $\alpha$, the optimization with respect to $\delta$ can be decomposed into a series of smaller optimization problems:
\begin{equation}
    \min_{\delta_i}\quad\frac{1}{2}(\alpha\cdot x_i-p_{i}+\delta_i)^2+\lambda|\delta_i|,\label{eq:obj-coord}
\end{equation}
for each $i=1,..,n_1$, which has a closed-form solution
\begin{equation}\label{delta_i}
    \delta_i^*=\mathrm{sign}(p_i-\alpha\cdot x_i)\max(0,|\alpha\cdot x_i-p_i|-\lambda).
\end{equation}
Plugging in the above into the objective \eqref{eq:obj-coord}, we can see that the objective is equal to:
\begin{equation}
    \frac{1}{2}(\alpha\cdot x_i-p_i+\delta_i)^2+\lambda|\delta_i|=f_{\mathrm{Huber}}(\alpha\cdot x_i-p_i;\lambda),
\end{equation}
\subsection{Proof of Theorem \ref{th1}}
Note that the subgradient of Huber loss is given by:
\begin{equation}
    \frac{\partial}{\partial z}f_{\mathrm{Huber}}(z;\lambda)=\begin{cases}
        z&|z|<\lambda\\
        -\lambda&z\leq-\lambda\\
        \lambda&z\geq \lambda
    \end{cases}
\end{equation}
Then, by the KKT conditions: 
\begin{equation}
\begin{split}
       \sum_{i\in\mathcal{I}_1}(\bar{\alpha}\cdot x_i-p_i)x_i-\lambda\left(\sum_{i\in\mathcal{I}_2}x_i-\sum_{i\in\mathcal{I}_3}x_i\right)=0.
\end{split}
\end{equation}
Hence, by taking the differentials of the above condition, we can obtain the closed-form solution \eqref{eq:dadlambda}.
% in the appendices.

\end{document}